\documentclass[twoside,titlepage,final,11pt]{article}
\usepackage{graphicx}
\usepackage{latexsym}
\usepackage{amssymb}
\usepackage{hyperref}

\catcode`\@=11
\def\ps@ppt{\def\@oddhead{\qquad \textsc{The Double Simplex}\hfil \thepage\qquad}
\def\@evenhead{\qquad\thepage \hfil \textsc{Chris Quigg} \qquad}
\def\@oddfoot{}\def\@evenfoot{}}    
\catcode`\@=12

\def\url#1{\mbox{\href{#1}{\sf #1}}}

\newcommand{\HRule}{\centerline{\rule{0.85\linewidth}{1mm}}}

\newcommand{\gev}{\hbox{ GeV}}

\newcommand{\tev}{\hbox{ TeV}}

\DeclareMathSymbol{\lll}{\mathrel}{AMSa}{"6E}

\newcommand{\cp}{\ensuremath{\mathcal{CP}}}

\def\ltap{\mathop{\raisebox{-.4ex}{\rlap{$\sim$}} 
\raisebox{.4ex}{$<$}}}

\newcommand{\cfrac}[2]{\textstyle \frac{#1}{#2}}

\newcommand{\m}{\hbox{ m}}

\newcommand{\smgg}{\ensuremath{\mathrm{SU(3)_c} \otimes \mathrm{SU(2)}_{\mathrm{L}} \otimes \mathrm{U(1)}_Y}}
\newcommand{\ewgg}{\ensuremath{\mathrm{SU(2)}_{\mathrm{L}} \otimes \mathrm{U(1)}_Y}}

\def\slashii#1{\setbox0=\hbox{$#1$}             
   \dimen0=\wd0                                 
   \setbox1=\hbox{\sl/} \dimen1=\wd1            
   \ifdim\dimen0>\dimen1                        
      \rlap{\hbox to \dimen0{\hfil\sl/\hfil}}   
      #1                                        
   \else                                        
      \rlap{\hbox to \dimen1{\hfil$#1$\hfil}}   
      \hbox{\sl/}                               
   \fi}                                         %
\def\slashiii#1{\setbox0=\hbox{$#1$}#1\hskip-\wd0\hbox to\wd0{\hss\sl/\/\hss}}

\begin{document}
\begin{flushright}
	\textsf{FERMILAB--CONF--05/371--T }
\end{flushright}
\vspace*{\stretch{1}}
\HRule
\begin{center}
	{\Huge The Double Simplex}\\[5mm]
	{\large Chris Quigg* } \\[5mm]
	Theoretical Physics Department \\
	Fermi National Accelerator Laboratory \\ P.O. Box 500 $\cdot$ Batavia, Illinois 60510 USA  \\[8mm]

	\centerline{\textsc{a new way to envision particles and interactions}}
	
	\vspace*{24pt}
	\parbox{3.3in}{
	Contribution to GustavoFest, a Symposium in
	Honor of Gustavo C.\ Branco, \textit{\cp\ Violation and the Flavor Puzzle,}
	Instituto Superior T\'{e}cnico,  Lisboa, Portugal $\cdot$ July 20,
	2005}
\end{center}
\HRule
\vspace*{\stretch{1.5}}
		*E-mail:quigg@fnal.gov.
\newpage
\setlength{\parindent}{2ex}
\setlength{\parskip}{12pt}
\section{Felicita\c{c}\~{o}es!}
It is my great pleasure to join in celebrating Gustavo Branco's
research in \cp\ violation and flavor physics, and to 
look forward to many  insightful contributions still to come. I am 
also very glad to have the opportunity to express my appreciation for 
the school of theoretical physics that has developed here in Lisbon 
through the efforts of Gustavo and colleagues. I particularly respect 
the attention to challenging problems that characterizes the 
Lisbon school, and I will allude to some of those hard problems in my 
remarks today.

\section{Presenting Particle Physics}
For some time, I have been concerned about the constricted portrayal of
the aspirations of particle physics in the popular scientific media.
Not long ago, when the first superconducting dipole was lowered into
the Large Hadron Collider tunnel at CERN, the BBC informed its
listeners that the point of the LHC is ``to discover the sought-after
Higgs boson, or `God particle,' which explains why matter has mass.''
An indistinct feeling that particle physics will be over, once we succeed in
ticking off this year's Holy Grail, is standard fare in the press.  Of
course, it is Quantum Chromodynamics (the strong interaction), not the
Higgs boson, that generates most of the visible mass in the Universe,
so even this straitened view of what the LHC may bring is garbled.

More troubling to me, reliance on shorthands such as ``the search for
the Higgs boson'' (as a token for uncovering the origin of electroweak
symmetry breaking through a thorough exploration of the 1-TeV scale)
seems to have narrowed discourse within our field.  It is not rare for
our own colleagues to talk in terms of a very limited menu of
opportunities for discovery.  That would be cause for concern, even if
the shorthands didn't spill over into the media.  As we embark upon the
LHC adventure, we will need open and prepared minds!

\begin{figure}[tb]
\begin{center}
\includegraphics[width=6.3cm]{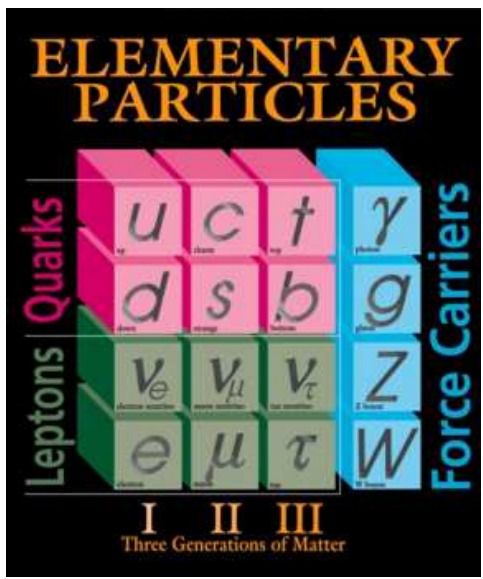}
\caption{The ubiquitous chart of quarks, leptons, and force carriers. \label{fig:smodel}}
\end{center}
\end{figure}

The iconic representations of the standard model of particle physics 
summarize the state of our knowledge, but conceal 
the state of our ignorance. I have a lot of affection for the table of 
particles and interactions shown in Figure~\ref{fig:smodel}, which 
represents the enthusiastic work of many teachers and grew out of a 
Conference on the Teaching Modern Physics held at Fermilab in 
1986. It encodes a lot of information in compact form, 
and---when used to start a conversation---can be a very attractive 
visual aid. But the chiseled-in-stone appearance makes it look so finished!

\begin{figure}[tbh]
\begin{center}
\includegraphics[width=10cm]{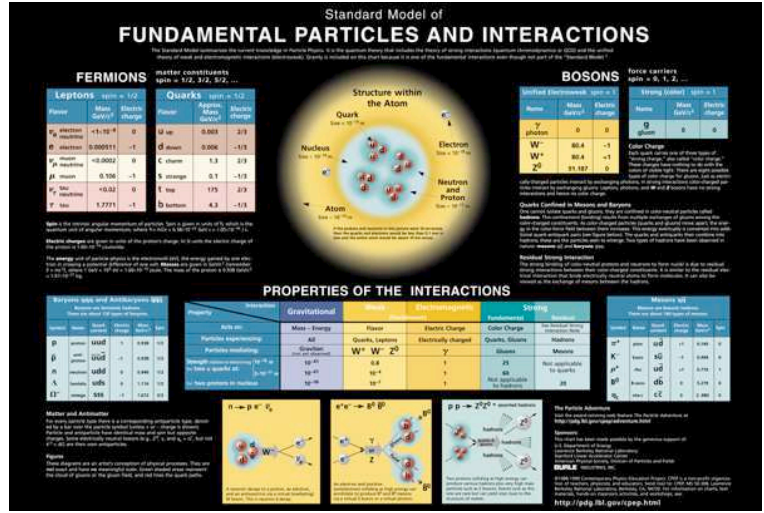}
\caption{CPEP's standard-model wall chart. 
\label{fig:wallchart}}
\end{center}
\end{figure}

The celebrated chart (see Figure~\ref{fig:wallchart}) of fundamental
particles and interactions created by the Contemporary Physics
Education Project
(\url{http://www.cpepweb.org/}), and
available as wall chart, poster, and place mat, has had a global 
reach---more than 200,000 distributed. It, too, has helped move 
particle physics into the classroom, and it presents many 
essential notions of the standard model.

Neither of these valuable pedagogical tools does much to suggest the 
plethora of open questions that animate current research in the 
field---questions to which we are led, in part, by the success of the 
standard model. As Presidential Science Advisor John Marburger 
pointedly observed to me recently, I have been telling him for two 
decades how important is the search for the agent of electroweak 
symmetry breaking, and yet the idea of the Higgs boson is nowhere to 
be found in either of our popular charts.

I want to describe today a work in progress, my attempt to create a
three-dimensional object that expresses a new way to envision the
particles and interactions.  My goal is to represent what we know is
true, what we hope might be true, and what we don't know---in other
terms, to show the connections that are firmly established, those we
believe must be there, and the open issues.

Galileo asserted that the Book of Nature is written in the language of 
mathematics. I am comfortable with the gentler claim that mathematics 
is one of the languages in which we have learned to listen best to 
Nature. I feel  that there is great value in striving to expose our insights 
and pose our new questions without hiding behind equations. I hope 
you will agree that the challenge of presenting our science without 
mathematical formalism is worthy and interesting. I hope, too, that 
you will be motivated to build on my approach or---even better---to 
make your own beginning.

Any chart or mnemonic device should be an invitation to narrative and a
spur to curiosity, and that is what I intend for the geometrical
construction I call the double simplex.  I want also to express the
spirit of play, of successive approximations, that animates the way
scientists work.  So instead of dissecting a completed double simplex, I will
build it up, step by step.  That will help us to see where choices are
to be made and to encounter some of the fascinating questions we face.

\section{Ground Rules \label{sec:gr}}
\begin{figure}[tb]
\begin{center}
\includegraphics[width=6.3cm]{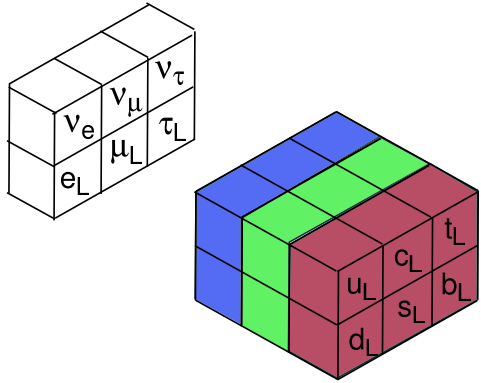}
\caption{The left-handed doublets of quarks and leptons that inspire 
the structure of the electroweak theory. \label{fig:DumbL}}
\end{center}
\end{figure}

Our picture of matter is based on the discovery of a set of pointlike
constituents: the quarks and the leptons, as depicted in
Figure~\ref{fig:DumbL},
and a few fundamental forces derived from gauge symmetries.  The quarks
are influenced by the strong interaction, and so carry \textit{color},
the strong-interaction charge, whereas the colorless leptons do not
feel the strong interaction.  
The pairs $(\nu_{e},e)$, $(u,d)$, etc., are emblematic of the 
transitions induced by the charged-current weak interactions.
The notion that the quarks and leptons
are elementary---structureless and indivisible---is necessarily
provisional, limited by our current resolution, $r \ltap
10^{-18}\m$.

\begin{figure}[tb]
\begin{center}
\includegraphics[width=10cm]{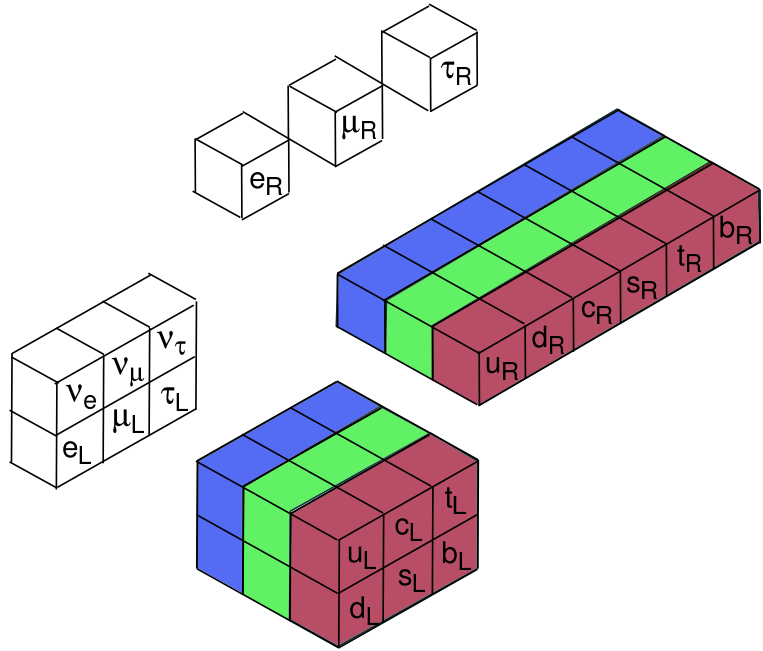}
\caption{Left-handed doublets, right-handed singlets of quarks 
and leptons.\label{fig:DumbSM}}
\end{center}
\end{figure}

Looking a little more closely at the constituents of matter, we find 
that our world is not as neat as the simple cartoon vision of 
Figure~\ref{fig:DumbL}. The left-handed and right-handed fermions 
behave very differently under the influence of the weak 
interactions. A more complete picture is given in 
Figure~\ref{fig:DumbSM},
which represents the way we looked at the world before the 
discovery of neutrino oscillations that require neutrino mass and 
almost surely imply the existence of right-handed 
neutrinos.\footnote{It is probably time to adopt the quark convention 
and label the neutrinos by mass eigenstates $\nu_{1},\nu_{2},\nu_{3}$ 
instead of the flavor eigenstates $\nu_{e},\nu_{\mu},\nu_{\tau}$.} Neutrinos 
aside, the striking fact is the asymmetry between left-handed fermion 
doublets and right-handed fermion singlets, which is manifested in 
\textit{parity violation} in the charged-current weak interactions. 
What does this distinction mean?

\begin{figure}[tb]
\begin{center}
\includegraphics[width=12cm]{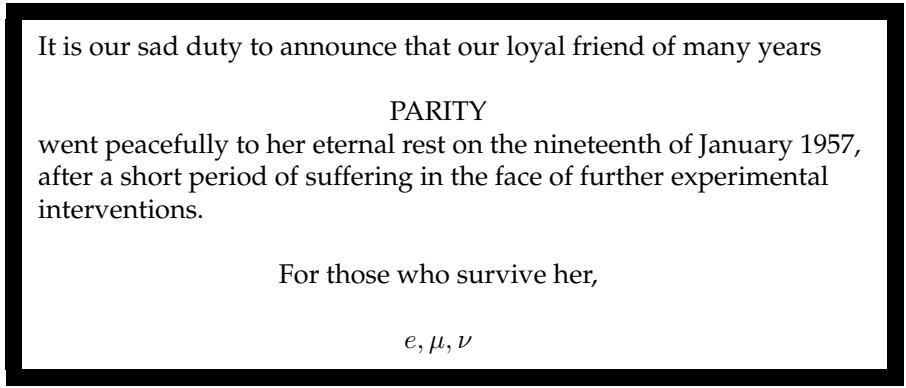}
\caption{Pauli to Weisskopf, on learning of parity 
violation in $\beta$ decay.\label{fig:Wolfi}}
\end{center}
\end{figure}

I think that everyone in this amphitheatre learned about parity
violation at school, and perhaps familiarity has dulled our senses a 
bit. [I estimate that I have personally written down the left-handed 
doublets $10^{4}$ times \ldots] It is worth remembering that parity 
violation came as a stunning surprise to our
scientific ancestors. Wolfgang Pauli was moved to send a 
black-bordered note to Victor Weisskopf, bearing the text shown in 
Figure~\ref{fig:Wolfi}.
It seems to me that nature's 
broken mirror---the distinction between left-handed and right-handed 
fermions---qualifies as one of the great mysteries. Even if we will 
not get to the bottom of this mystery next week or next year, it 
should be prominent in our consciousness---and among the goals we 
present to others as the aspirations of our science.

The insight that local gauge symmetries imply interactions leads us to 
see the $W^{\pm}$ bosons as the agents of symmetry transformations 
that take $u \leftrightarrow d$, and so forth. In similar fashion, the 
gluons of SU(3) are symmetry operators that transform quark 
colors, preserving flavor. But there is more to our understanding of the 
world than is revealed by identifying the symmetries represented in
Figure~\ref{fig:DumbSM}. The electroweak gauge symmetry is hidden,
$\ewgg \rightarrow \mathrm{U(1)_{em}}.$ If it were not, the world 
would be very different. 

At first sight, $\Box$~All the quarks and leptons would be massless and
move at the speed of light.  $\Box$ Electromagnetism as we know it
would not exist, but there would be a long-range weak-hypercharge force.
$\Box$  QCD would confine quarks and generate
baryon masses roughly as we know them.  $\Box$ The Bohr radius of
``atoms'' consisting of an electron or neutrino attracted by the
weak-hypercharge interaction to the nucleons would be infinite.  $\Box$ Beta
decay, inhibited in our world by the great mass of the $W$ boson, would
not be weak.  $\Box$ The unbroken $\mathrm{SU(2)}_{\mathrm{L}}$
interaction would confine objects that carry weak isospin.

It is fair to say that electroweak symmetry breaking shapes our world!
In fact, when we take into account every aspect of the influence of the
strong interactions, the analysis of how the world would be is very
subtle and fascinating.  Please take time to think about a 
not-so-simple homework problem: \textit{What would the
everyday world be like if there were {no mechanism, like the Higgs
mechanism, to break the electroweak symmetry?}  Consider the effects of
all of the \smgg\ gauge interactions.}

Because one of my goals is to devise a metaphor that evokes open 
questions, highlighting what we do not know, my intention is not to 
make predictions that prejudge Nature's answers, but to create a 
language nimble enough to relate many stories. A discovery that breaks 
out of the framework should represent not just a choice among 
alternatives we already recognize, but a fundamental revision of our 
thinking. After this prologue, let us begin!

\section{Toward the double simplex \label{sec:DS}}
Both quarks and leptons are spin-$\cfrac{1}{2}$,
pointlike fermions that occur in $\mathrm{SU(2)}_{\mathrm{L}}$ doublets.  The
obvious difference is that quarks carry $\mathrm{SU(3)_{c}}$ color
charge whereas leptons do not, so we could imagine that quarks and
leptons are simply distinct and unrelated species.  But we have reason
to think otherwise.  The proton's electric charge very closely
balances the electron's, $(Q_{p}+Q_{e})/e < 10^{-21}$,
suggesting that there must be a link between protons---hence,
quarks---and electrons---hence, leptons.  Moreover, quarks and leptons are
required, in matched pairs, for the electroweak theory to be 
anomaly-free, so that quantum corrections respect the symmetries on 
which the theory is based.

\begin{figure}[tbh]
\begin{center}
\includegraphics[width=4cm]{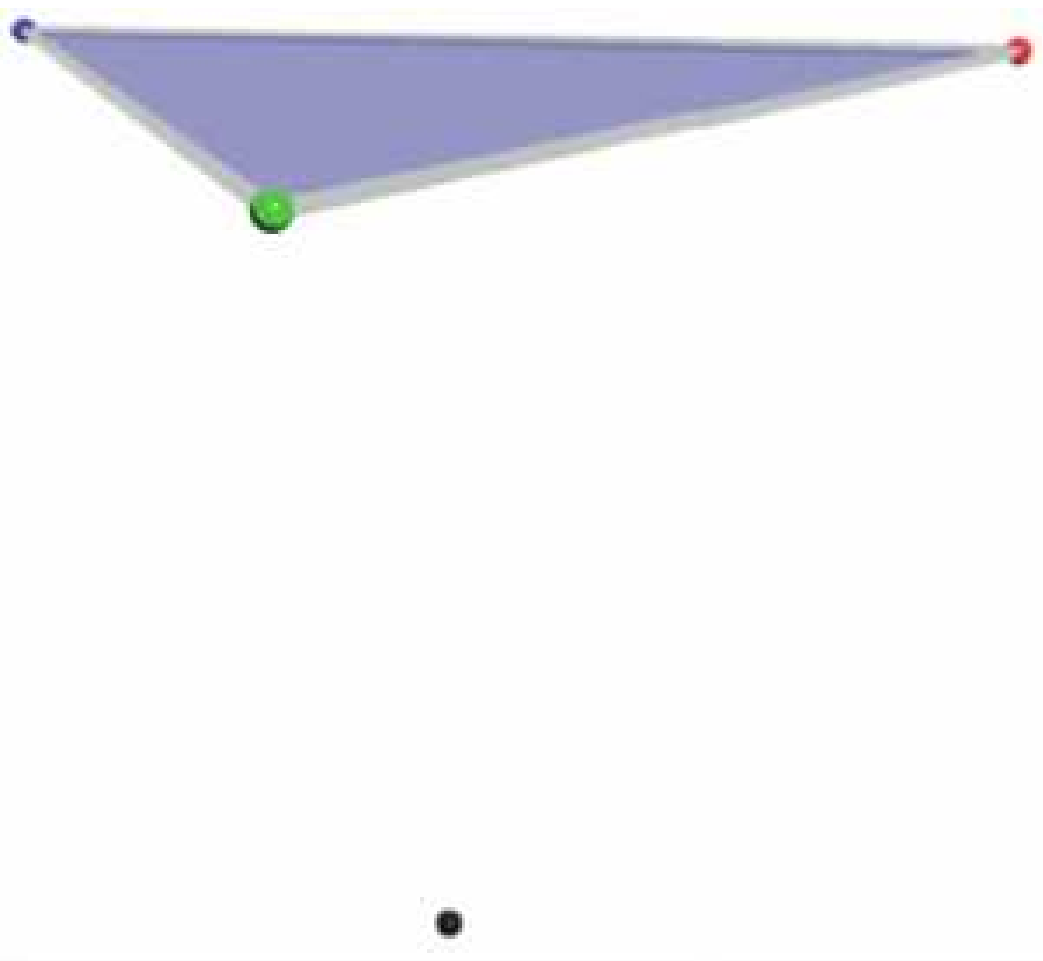}\qquad\qquad\qquad
\includegraphics[width=4cm]{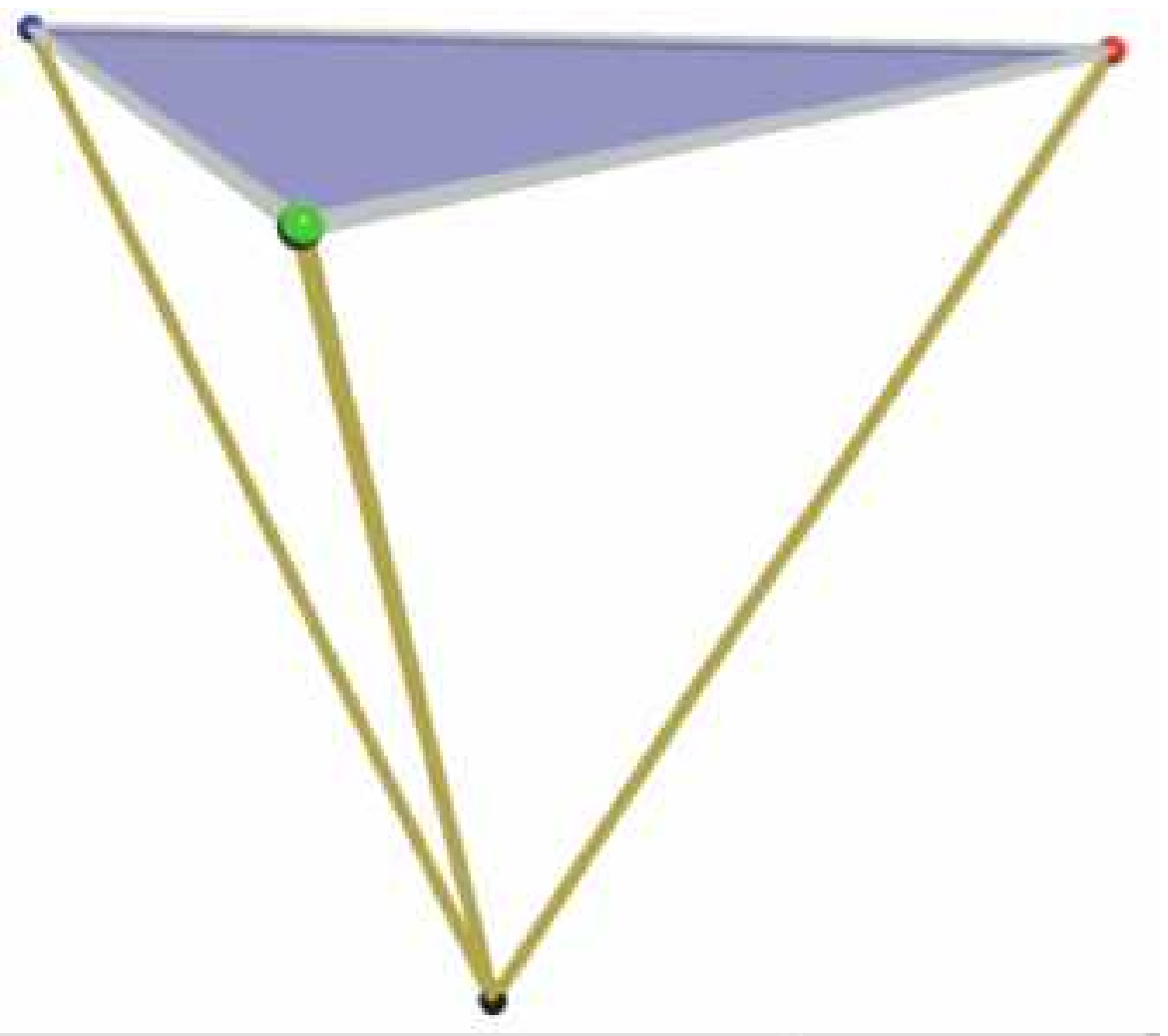}
\caption{Left panel: A color triplet of quarks and a color singlet lepton, 
arrayed to explore lepton number as a fourth 
color. Right panel: Tetrahedron representing the \textbf{4} representation of 
$\mathrm{SU(4)}$, showing the hypothetical leptoquark transitions. 
\label{fig:LLQ}}
\end{center}
\end{figure}

It is fruitful to display the color-triplet red, green, and blue quarks 
in the equilateral triangle weight diagram for the \textbf{3} 
representation of $\mathrm{SU(3)}_{\mathrm{c}}$, as shown in the left 
panel of 
Figure~\ref{fig:LLQ}. There I have filled in the plane between them to 
indicate the transitions mediated by gluons.
The equality of proton and (anti)electron charges and the need to 
cancel anomalies in the electroweak theory suggest that we join the 
quarks and leptons in an extended family, or multiplet. Pati and 
Salam provided an apt metaphor when they proposed that we 
regard lepton number as a fourth color. To explore that possibility, 
I have placed the lepton in Figure~\ref{fig:LLQ} at the apex of a 
tetrahedron that corresponds to the fundamental \textbf{4} 
representation of $\mathrm{SU(4)}$.

If $\mathrm{SU(4)}$ is not merely a useful \textit{classification 
symmetry} for the quarks and leptons, but a \textit{gauge symmetry,} 
then there must be new interactions that transform quarks into 
leptons, as indicated by the gold lines in the right panel of 
Figure~\ref{fig:LLQ}. 
Are they present in Nature?
If leptoquark transitions do exist, they can mediate reactions that 
change baryon and lepton number, such as proton decay. The long 
proton lifetime tells us that  leptoquark transitions 
 must be far weaker than the strong, weak, and 
electromagnetic interactions of the standard model. What accounts for 
the feebleness of leptoquark transitions?

\begin{figure}[b!]
\begin{center}
    \includegraphics[width=4.7cm]{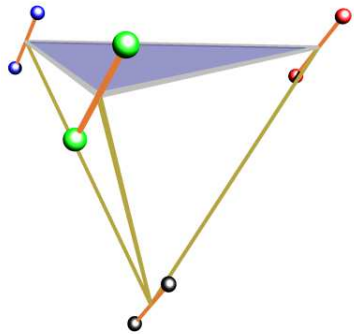}
\caption{The $\mathrm{SU(4)}$ tetrahedron, decorated with left-handed 
fermions.  \label{fig:LHdec}}
\end{center}
\end{figure}

Our world isn't built of a single quark flavor and a single lepton 
flavor. The left-handed quark and lepton doublets offer a key clue to 
the structure of the weak interactions. We can represent the $(u_{\mathrm{L}}, 
d_{\mathrm{L}})$ and $(\nu_{\mathrm{L}}, e_{\mathrm{L}})$ doublets by decorating the tetrahedron, as 
shown in Figure~\ref{fig:LHdec}. 
The orange stalks connecting $u_{\mathrm{L}} \leftrightarrow d_{\mathrm{L}}$ 
and $\nu_{\mathrm{L}} 
\leftrightarrow e_{\mathrm{L}}$ stand for the $W$~bosons that 
mediate the charged-current weak interactions. Using the same object 
for quark and lepton transitions is a token for the universality 
of the weak charged-current couplings.

What about the right-handed fermions? In quantum field theory, it is 
equivalent to talk about left-handed \textit{antifermions.} That 
observation motivates me to display the right-handed quarks and 
leptons as decorations on an inverted tetrahedron. 
\begin{figure}[tb]
\begin{center}
\includegraphics[width=4.6cm]{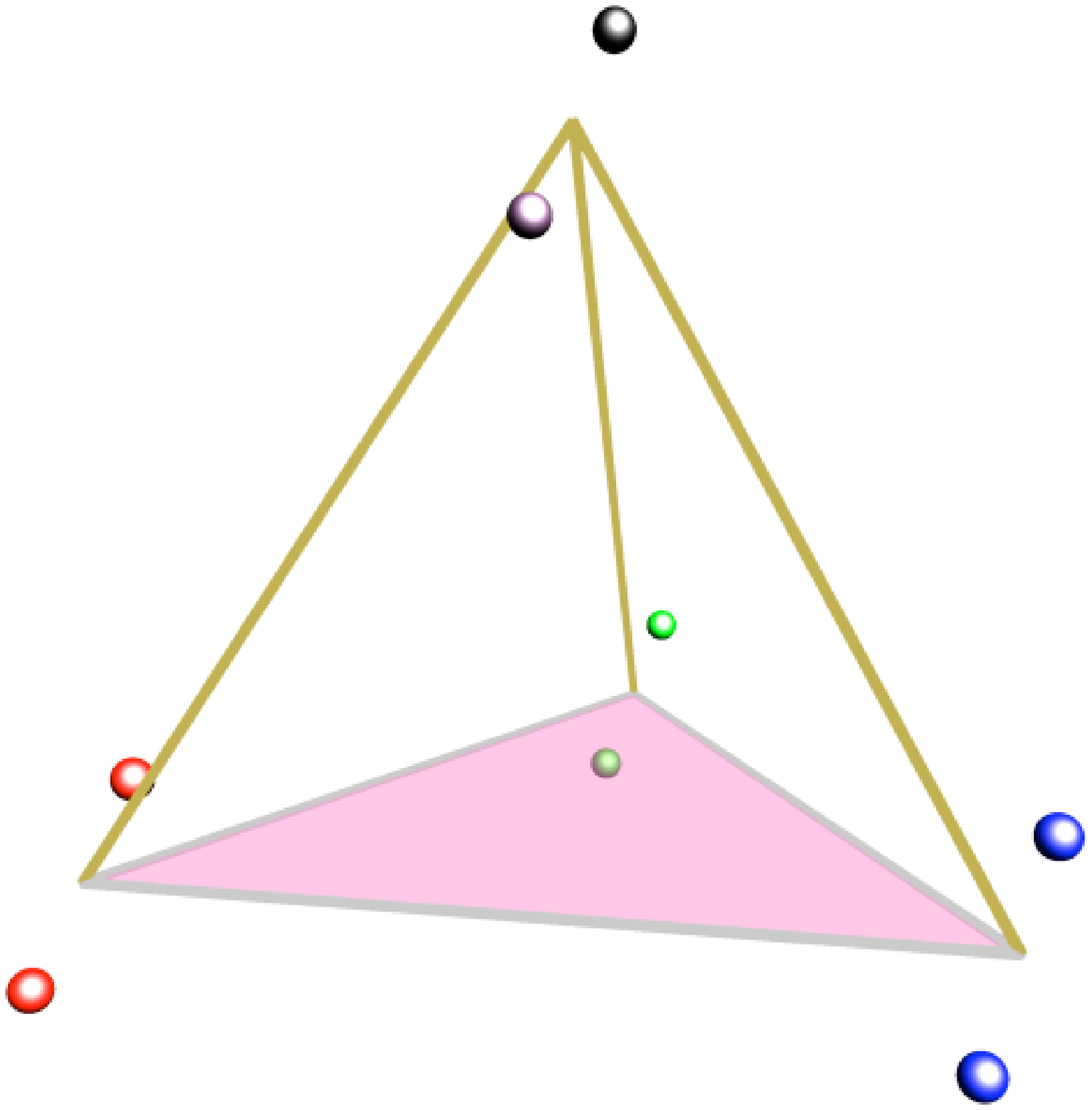}
\qquad\qquad
\includegraphics[width=4.6cm]{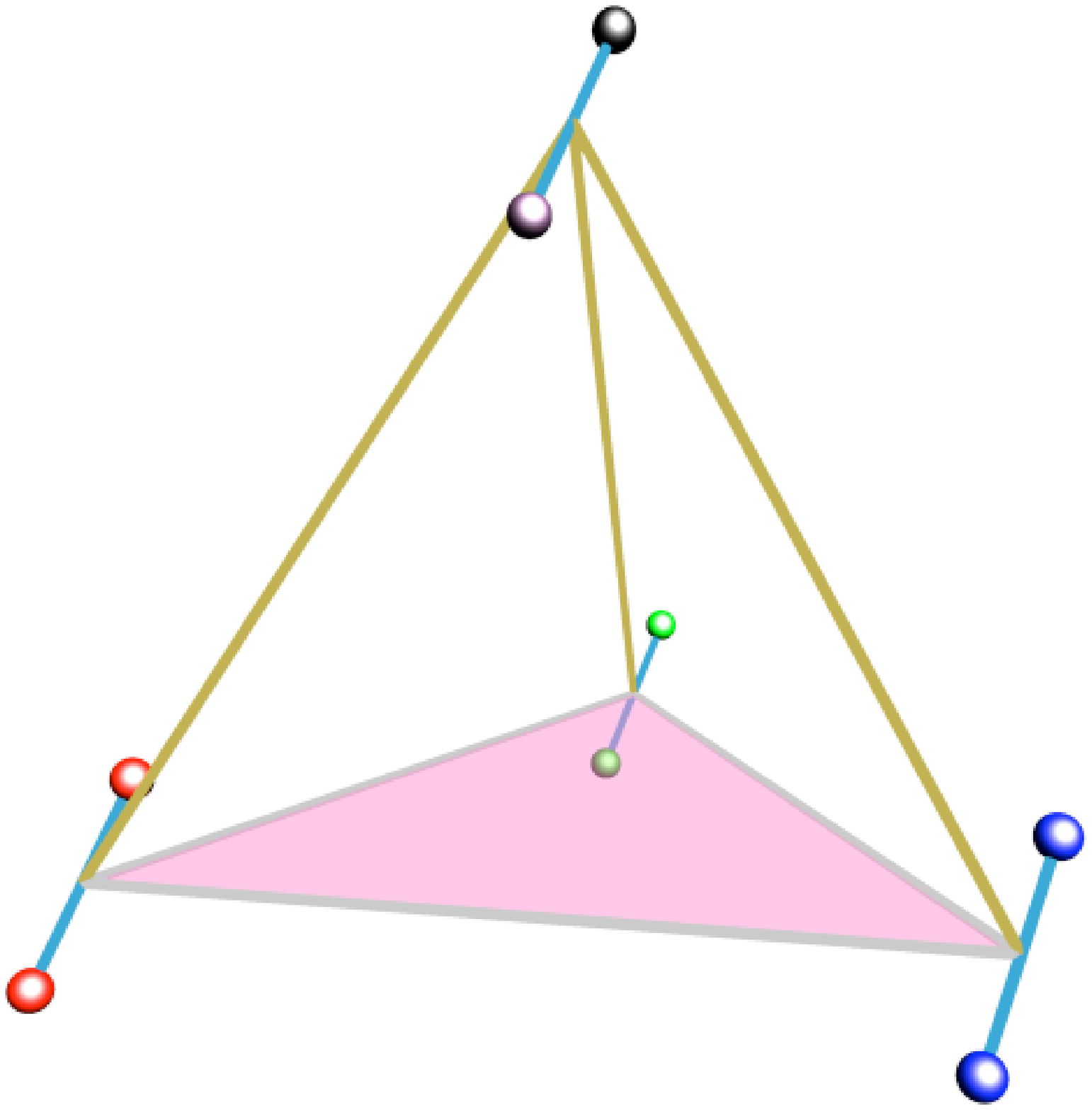}
\caption{The inverted tetrahedron, decorated with right-handed 
quark $(d_{\mathrm{R}},u_{\mathrm{R}})$ and lepton $(e_{\mathrm{R}}, \nu_{\mathrm{R}})$ pairs. The left 
panel depicts our current understanding, without right-handed charged 
currents; the right panel shows how a right-handed $W_{\mathrm{R}}$-boson could be 
added.\label{fig:RHdec}}
\end{center}
\end{figure}
The right-handed 
fermions are, by definition, singlets under the usual left-handed weak 
isospin, $\mathrm{SU(2)}_{\mathrm{L}}$, so I give the decorations an 
orthogonal orientation. 
Neutrino oscillations make us almost certain that a right-handed 
neutrino exists,\footnote{A purely left-handed Majorana mass term 
remains a logical, though not especially likely, possibility.} so I have placed a 
right-handed neutrino in Figure~\ref{fig:RHdec}. I have given it a 
different coloration from the established leptons as a reminder that we 
have not proved its existence, and we do not know its nature.

We do not know whether the pairs of quarks and leptons 
carry a right-handed weak isospin, in other words, whether they make 
up $\mathrm{SU(2)}_{\mathrm{R}}$ doublets. We do know that we have---as 
yet---no experimental evidence for right-handed charged-current weak 
interactions. Accordingly, I will generally display the right-handed fermions 
without a connecting $W_{\mathrm{R}}$-boson, as shown in the left panel of 
Figure~\ref{fig:RHdec}.
Is there a right-handed charged-current interaction? If not, we come 
back to the question that shook our ancestors: what is the meaning of 
parity violation, and what does it tell us about the world? If we 
should discover---or wish to conjecture---a right-handed charged 
current, it can be added to our graphic, as shown in the right-panel 
of Figure~\ref{fig:RHdec}. If there is a right-handed charged-current 
interaction, restoring parity invariance at high energy scales, what 
makes that interaction so feeble that we haven't yet observed it?

\begin{figure}[tb]
\begin{center}
\raisebox{10pt}{\includegraphics[width=5.2cm]{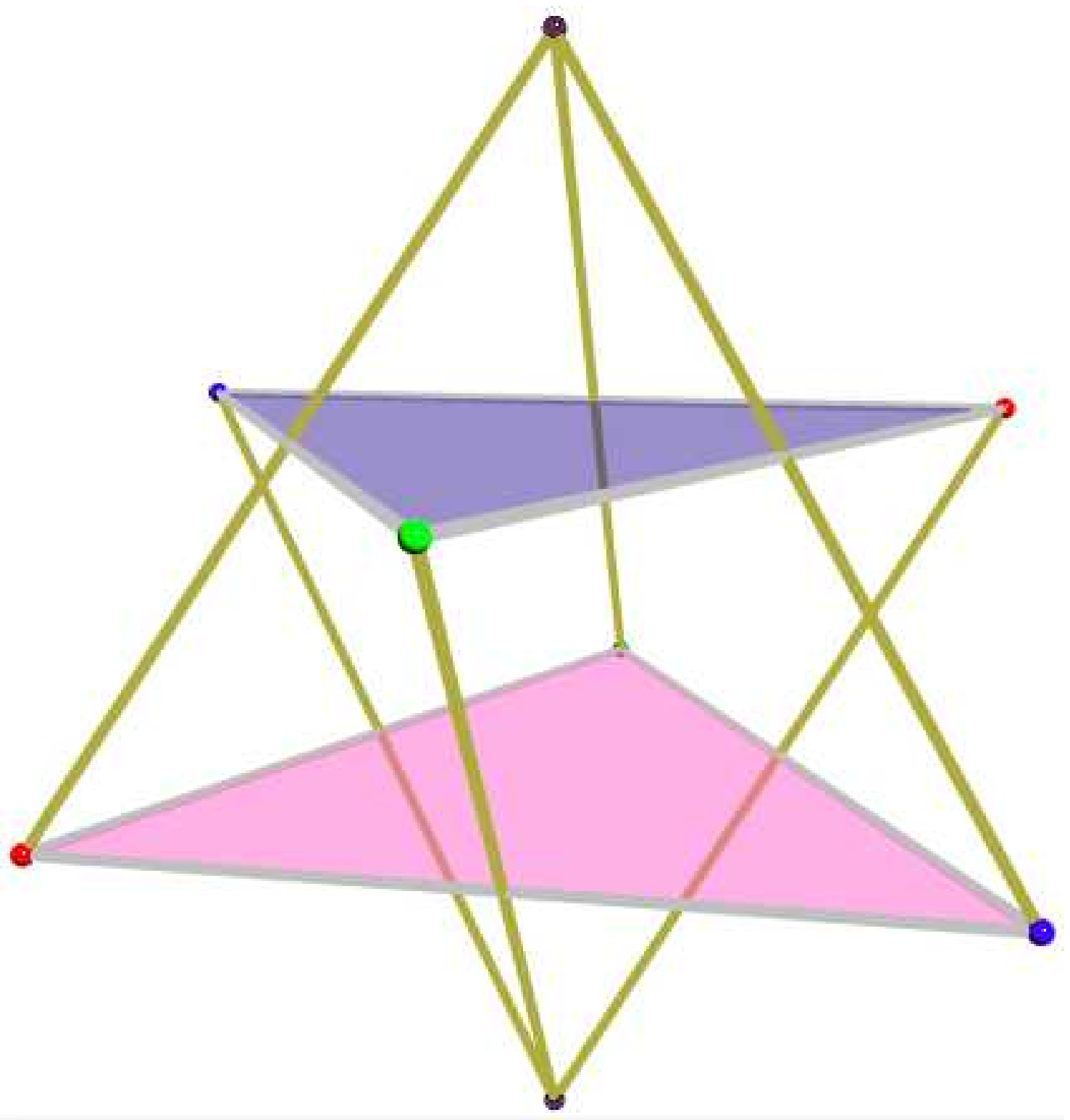}}
\qquad
\includegraphics[width=6.0cm]{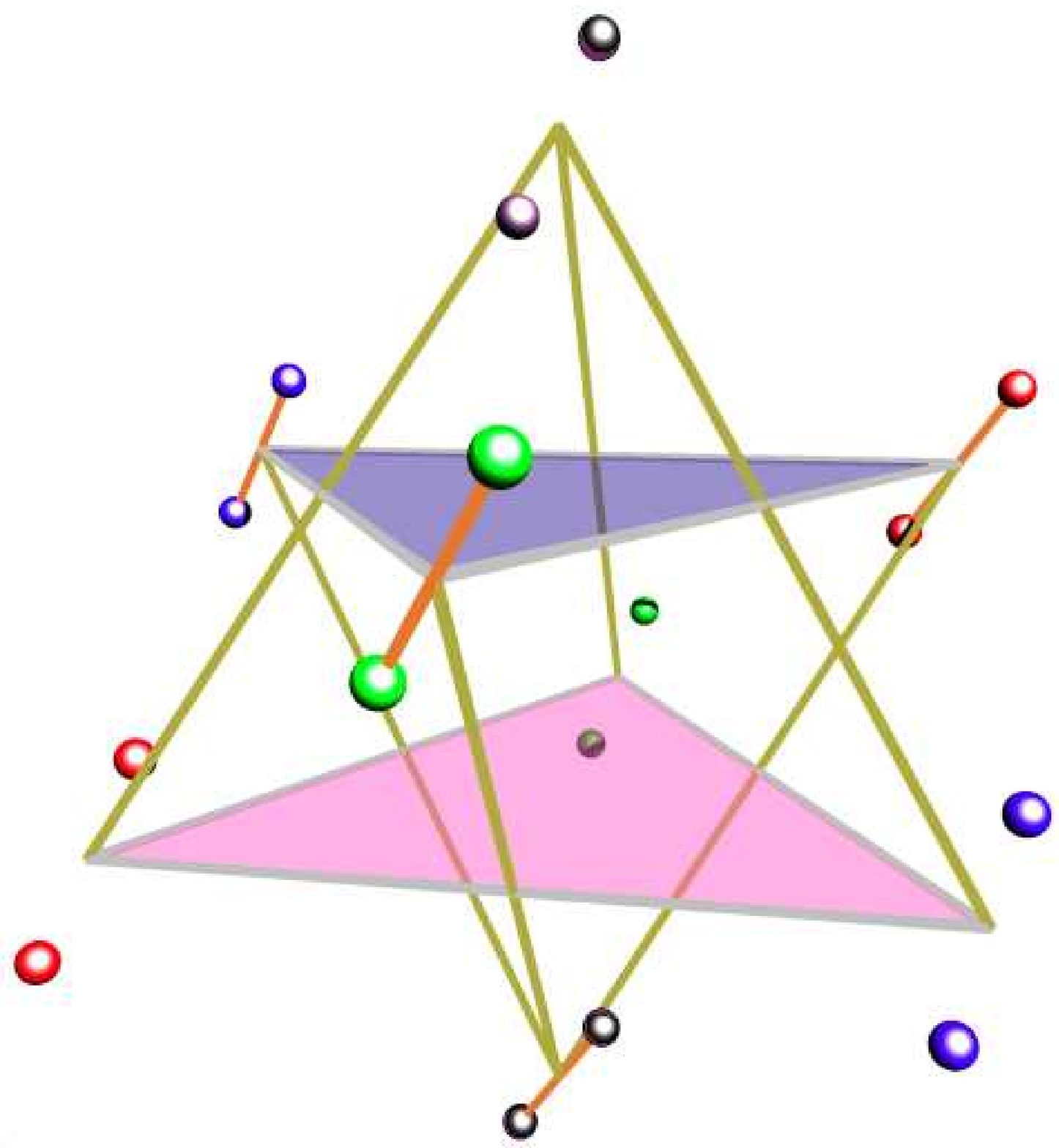}
\caption{The double simplex, undecorated (left panel) and decorated
with one generation of quarks and leptons (right panel).  \label{fig:onegenDS}}
\end{center}
\end{figure}

If parity violation in the weak interactions teaches us of an 
important asymmetry between left-handed and right-handed fermions, the 
nonvanishing masses of the quarks and leptons inform us that left and 
right cannot be entirely separate. Coupling the left-handed particle 
to its right-handed counterpart is what endows fermions with mass. 
For example, the mass term of the electron in the Lagrangian of 
quantum electrodynamics is 
\begin{equation}
    \mathcal{L}_{e} = -m_{e}\bar{e}e = 
-m_{e}\bar{e}\left[\cfrac{1}{2}(1-\gamma_{5}) + 
\cfrac{1}{2}(1+\gamma_{5})\right]e =
-m_{e}(\bar{e}_{\mathrm{R}}e_{\mathrm{L}} + 
\bar{e}_{\mathrm{L}}e_{\mathrm{R}})\;.
\label{eq:emass}
\end{equation}
How shall we combine left with right?  A suggestive structure is the 
pair of interpenetrating tetrahedra shown in 
Figure~\ref{fig:onegenDS}.
Mathematicians refer to a tetrahedron as a \textit{simplex} in 
three-dimensional space, so I call this construction the \textit{double 
simplex.} We will return to the question of mass momentarily.

\begin{figure}[tb]
\begin{center}
\raisebox{13pt}{\includegraphics[width=5.2cm]{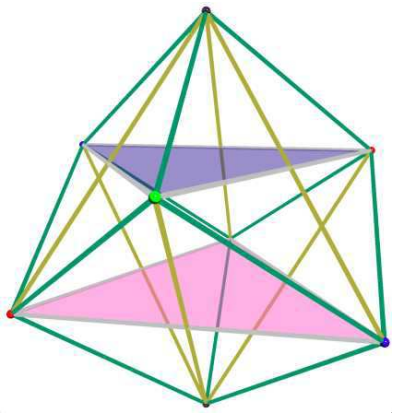}}
\caption{The double simplex, with additional interactions suggested by 
the shape of the figure indicated as green links. \label{fig:cube}}
\end{center}
\end{figure}

Now think of holding the double simplex in your hand, turning it over
to behold its symmetrical form.  Your eyes may be drawn to link the
nearest unconnected vertices, as shown in Figure~\ref{fig:cube}.
When that happens, you are visualizing one aspect of unification: When
we combine the two sets of particles into one representation, we are
invited to consider the possibility of new transformations that take
any member of the extended family into any other.  In this case, the
hypothetical new interactions are easy to visualize, because the double
simplex can be inscribed in a cube.  Do some of these interactions
exist?  If so, why are they so weak that we have not yet observed them?

If we think of the double simplex as composed of left-handed particles
and left-handed antiparticles, the agents of change will be new gauge
bosons, since gauge-boson interactions preserve chirality. If we take 
the two tetrahedra to stand for left-handed and right-handed 
particles, then the new connections will be spin-zero particles.

Fermion masses tell us that the left-handed and right-handed fermions 
are linked, but we do not know what agent makes the connection. In the 
standard \ewgg\ electroweak theory, it is the Higgs boson---the avatar 
of electroweak symmetry breaking---that endows the fermions with mass. 
But this has not been proved by experiment, and it is certainly 
conceivable that some entirely different mechanism is the source of 
fermion mass.

\begin{figure}[tb]
\begin{center}
\includegraphics[width=6cm]{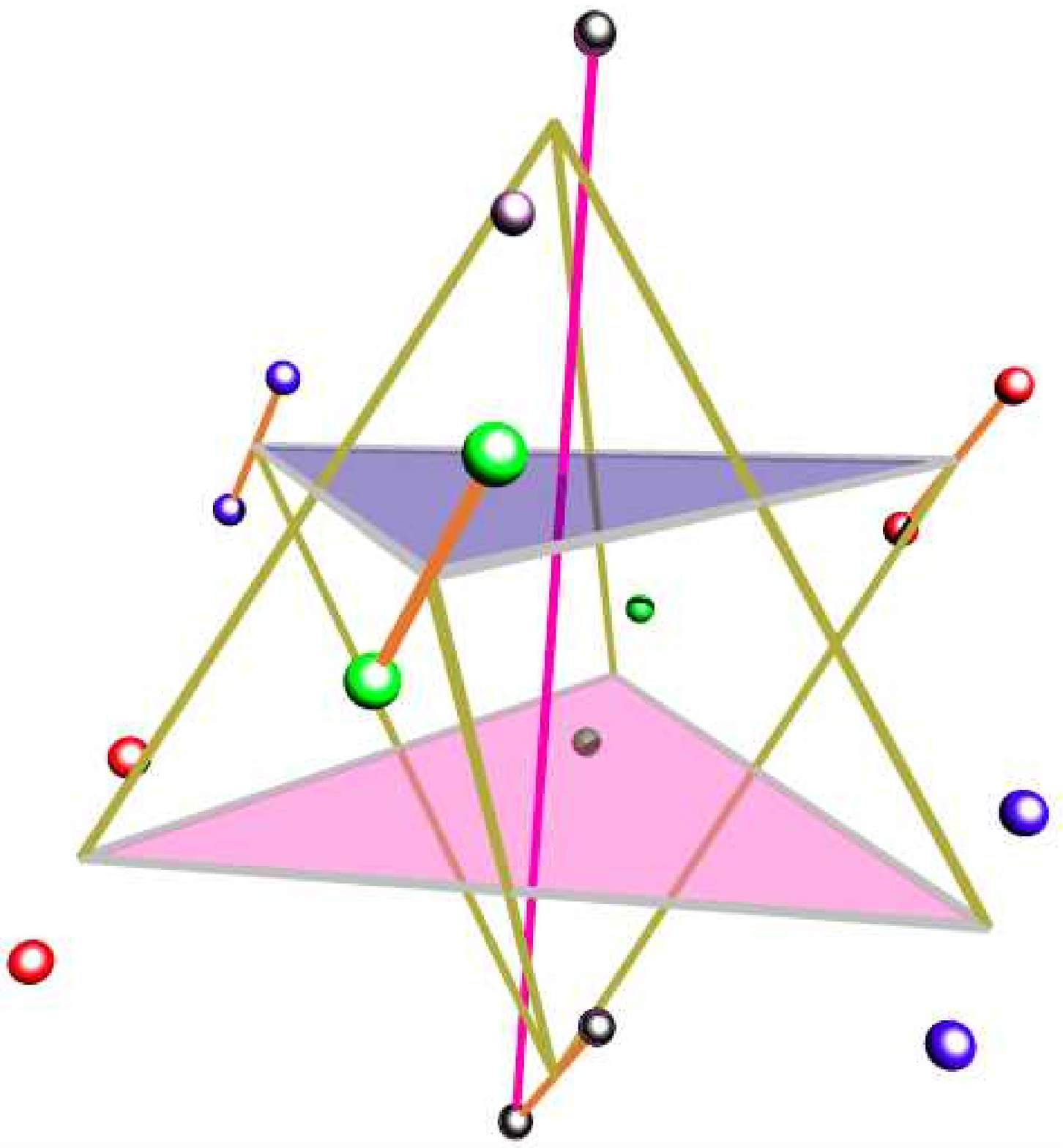}
\phantom{III}
\includegraphics[width=6cm]{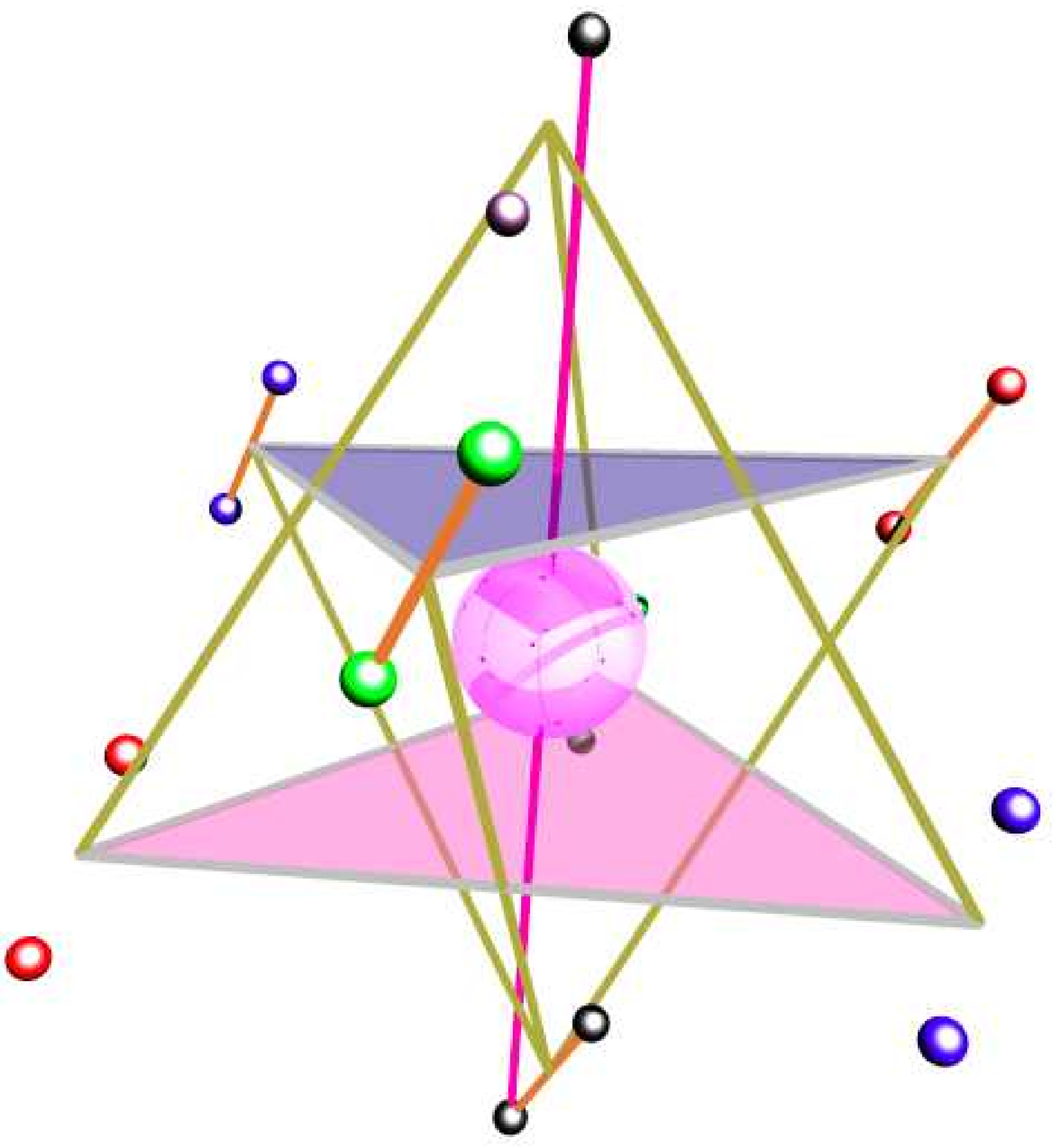}
\caption{The connection between $e_{\mathrm{R}}$ and $e_{\mathrm{L}}$
implied by the electron's nonzero mass.  \label{fig:onegene}}
\end{center}
\end{figure}

I draw the connection between the left-handed and right-handed
electrons in Figure~\ref{fig:onegene}.  
The left-hand panel shows the link between $e_{\mathrm{L}}$ and
$e_{\mathrm{R}}$.  In the right-hand panel, I show the connection
veiled within an iridescent globe that represents our ignorance of the
symmetry-hiding phase transition that links left and right. 
[Critical opalescence is a marker for a phase transition.] It is 
excellent to find that the central mystery of the standard model---the 
nature of electroweak symmetry breaking---appears at the center of the 
double simplex!

\begin{figure}[tb]
\begin{center}
\includegraphics[width=6cm]{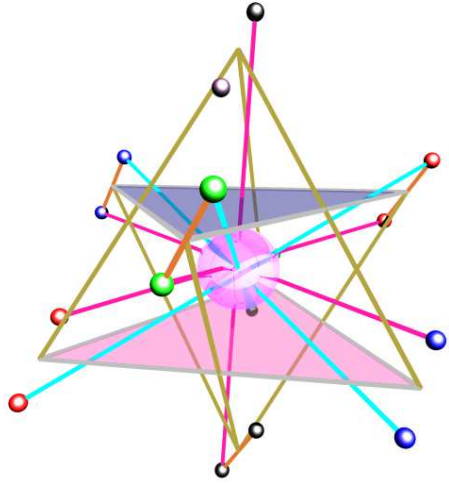}
\caption{The connections that give rise to mass for the quarks and 
leptons of the first generation. \label{fig:onegen}}
\end{center}
\end{figure}

Connecting all the left-handed fermions to their right-handed 
counterparts\footnote{I omit the neutrinos in this brief tour, because 
there are several possible origins for neutrino mass.} leads us to the 
representation given in Figure~\ref{fig:onegen}.
Does one agent give masses to all the quarks and leptons? (That is 
the standard-model solution.) If so, what distinguishes one fermion 
species from another? We do not know the answer, and for that reason I 
contend that \textit{fermion mass is evidence for physics beyond the standard 
model.} Let us illustrate the point in the standard-model context. The 
 mass of fermion $f$ is given by 
\begin{equation}
    \mathcal{L}_{f} =
-\displaystyle{\frac{{}{\zeta_{f}}v}{\sqrt{2}}}(\bar{f}_{\mathrm{R}}f_{\mathrm{L}} +
\bar{f}_{\mathrm{L}}f_{\mathrm{R}}) =
-\displaystyle{\frac{{}{\zeta_{f}}v}{\sqrt{2}}}\bar{f}f\;,
\label{eq:fmassYuk}
\end{equation}
where
$v/\sqrt{2} = 
    (G_{\mathrm{F}}\sqrt{8})^{-1/2} \approx 174\gev$
is the vacuum expectation value of the Higgs field. The 
\textit{Yukawa coupling} $\zeta_{f}$ is not predicted by the 
electroweak theory, nor does the standard model relate different
 Yukawa couplings. In any 
event, we do not know whether one agent, or two, or three, will give rise 
to the electron, up-quark, and down-quark masses. 

Of course, the world we have discovered until now consists not only 
of one family of quarks and one family of leptons, but of the three 
pairs of quarks and three pairs of leptons. We do not know the 
meaning of the replicated generations, and indeed we have no 
experimental indication to tell us which pair of quarks is to be 
associated with which pair of leptons. 

\begin{figure}[tb]
\begin{center}
\includegraphics[width=6.3cm]{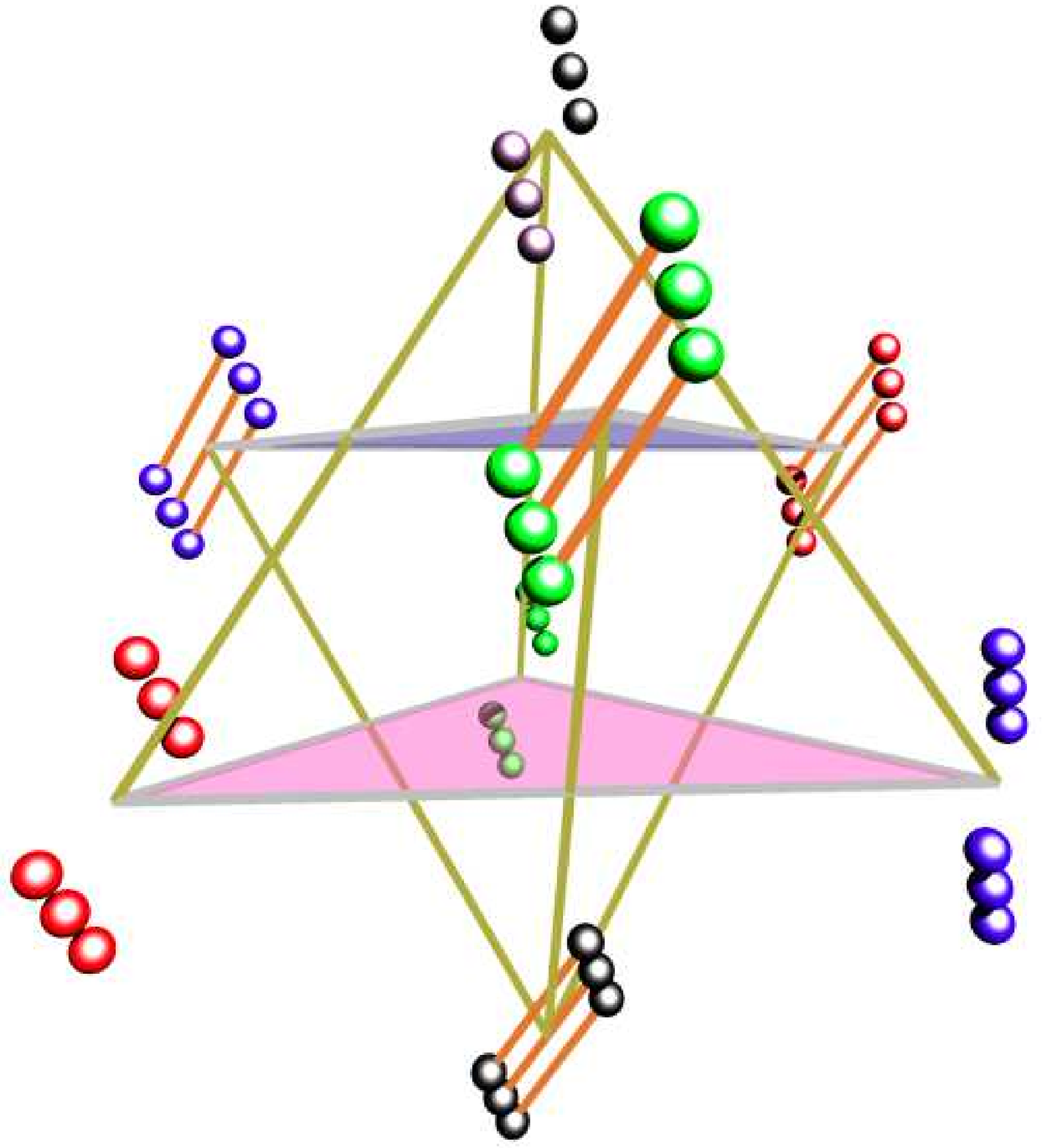}
\phantom{III}\includegraphics[width=6.3cm]{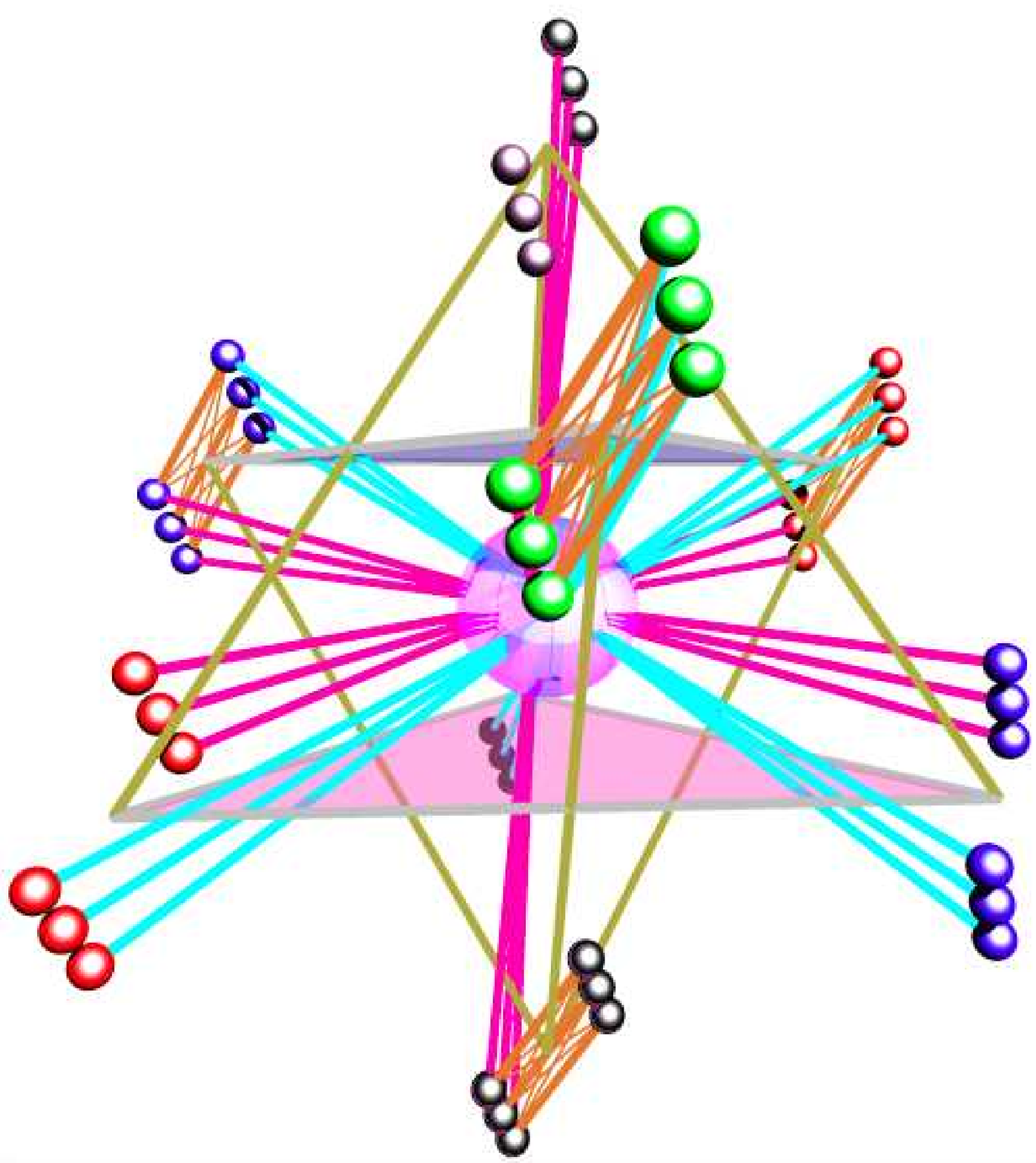}
\caption{Left panel: Three generations of quarks and leptons.  Right
panel: The connections that give rise to mass and mixing for three
generations of quarks and leptons.  \label{fig:threegen}}
\end{center}
\end{figure}

Lacking any understanding of the relation of one generation 
to another, I depict the three generations in the double simplex 
simply by replicating the decorations to include three pairs of 
quarks and three pairs of leptons, as shown in the left panel of 
Figure~\ref{fig:threegen}.
The connections that generate the fermion masses are indicated in the
right panel of Figure~\ref{fig:threegen}.  Look closely at the lines 
representing weak charged-current transitions, before and after left 
is joined with right. In the case of
more than one generation, the connections that endow the fermions with
mass also determine the mixing among generations, the suppressed
transitions such as $u \leftrightarrow s$ and $u \leftrightarrow b$.

We may not yet understand the character of the connections between 
left-handed and right-handed fermions, but this representation 
emphasizes that the connections are there. If, for example, you are a 
BaBar graduate student measuring a rare $B$ decay in order to 
determine the quark-mixing matrix parameter $V_{td}$, the double 
simplex shows that you are not measuring an isolated quantity, but 
one tied to many other aspects of particle physics.

With three generations, the Yukawa couplings may have complex phases
that give rise to $\mathcal{CP}$-violating transitions.  Although it is
correct to say that the standard model describes the observed examples
of $\mathcal{CP}$ violation, I would like to insist that because the
standard model does not prescribe the Yukawa couplings, $\mathcal{CP}$
\textit{violation---like fermion mass---is evidence for physics beyond
the standard model.}

\begin{figure}[tb]
\begin{center}
\includegraphics[width=6.3cm]{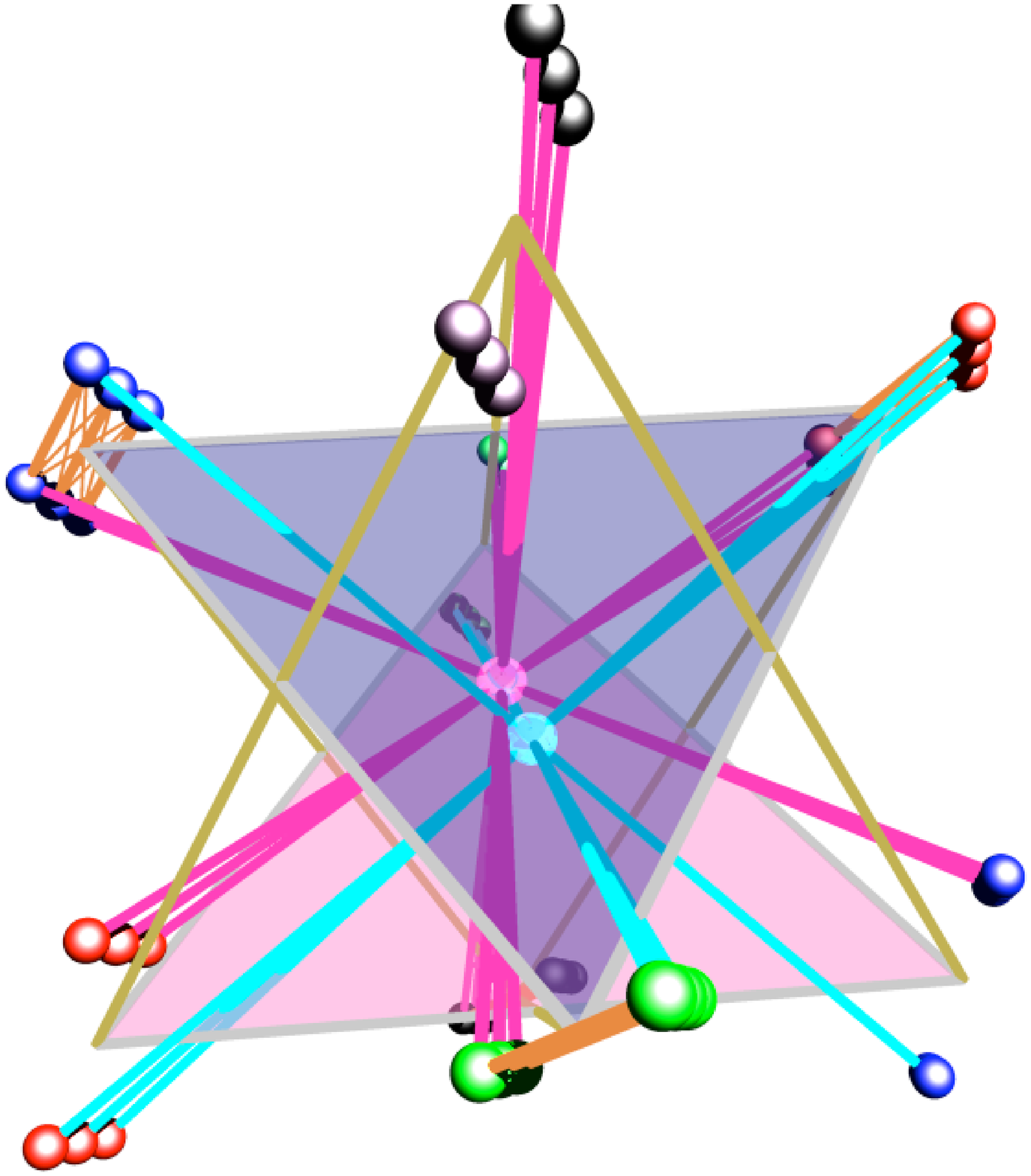}
\phantom{III}\includegraphics[width=6.3cm]{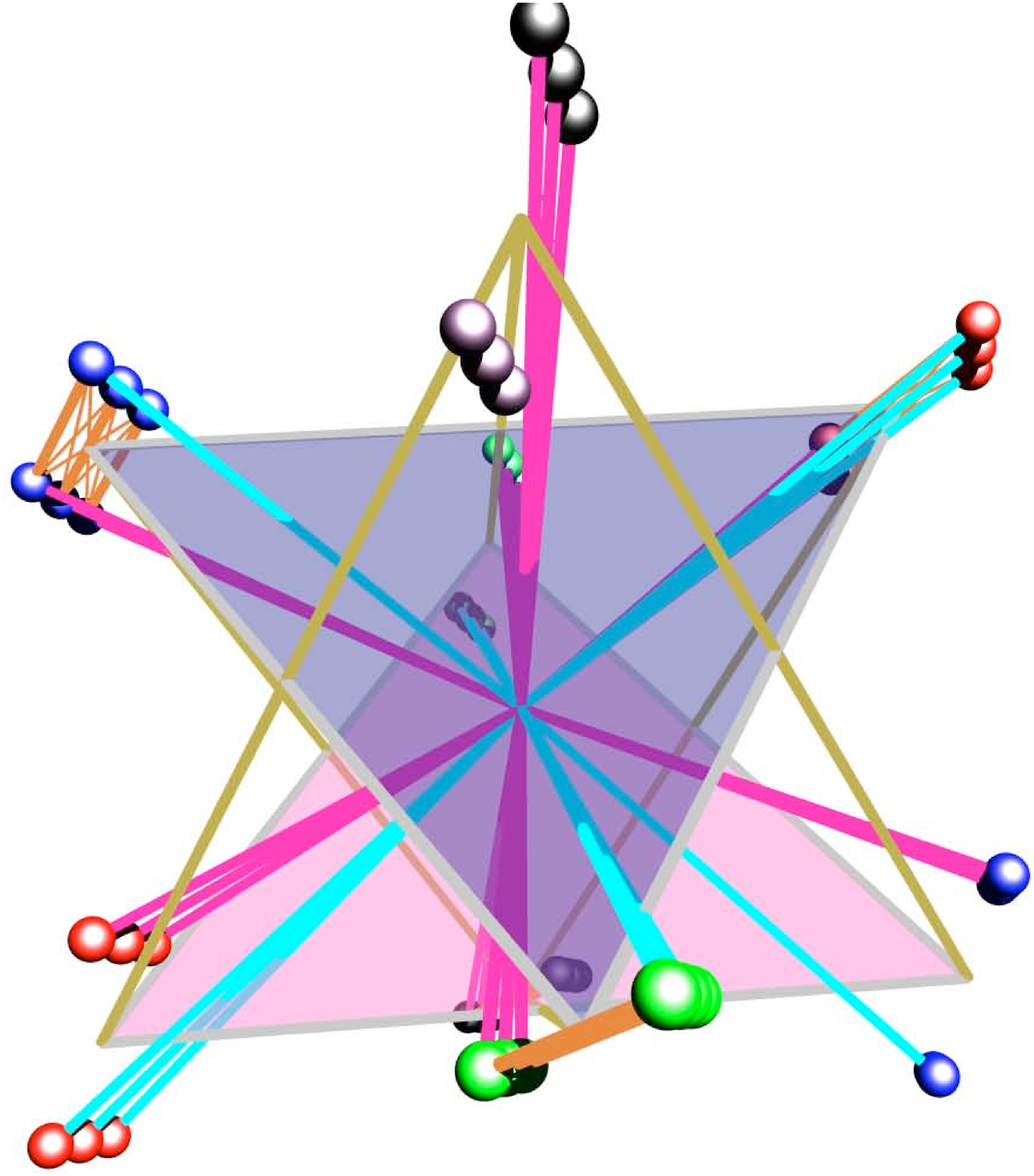}
\caption{Left panel: Connections between left-handed fermions and 
their right-handed counterparts, with separate agents for the 
$I_{3}=+\cfrac{1}{2}$ and $I_{3} = - \cfrac{1}{2}$ particles.  Right
panel: Standard-model depiction of the connections.  \label{fig:threegenopen}}
\end{center}
\end{figure}

I have said that I intend the double simplex as a device for 
eliciting questions. Might it also lead us to inventions? Here we 
have the first example. If we remove the iridescent globe to see what 
lies within, the left panel of Figure~\ref{fig:threegenopen} shows 
that all the $u,c,t$ connections pass through a common point, whereas 
the $d,s,b$, and charged-lepton connections pass through a distinct 
point. One agent gives mass to the weak-isospin $+\cfrac{1}{2}$ 
fermions, another agent sets the masses of the weak-isospin 
$-\cfrac{1}{2}$ fermions.

Now, this situation is not astounding to a roomful of theoretical 
physicists---we recognize is as characteristic of supersymmetric 
models or two-Higgs-doublet models. The fun comes because we did not 
(consciously!) build it in; we labeled the fermions in an obvious way 
and connected the dots. Had we never encountered the 
two-Higgs-doublet possibility before, we could have noticed it 
through the double simplex!

I hasten to add that this solution is not forced on us. The right 
panel of Figure~\ref{fig:threegenopen} represents the standard model, 
and it is easy to represent many other possibilities as well. As we 
hoped, the double simplex offers both a flexible language and a path 
to discovery.

\begin{figure}[tb]
\begin{center}
\includegraphics[width=7.5cm]{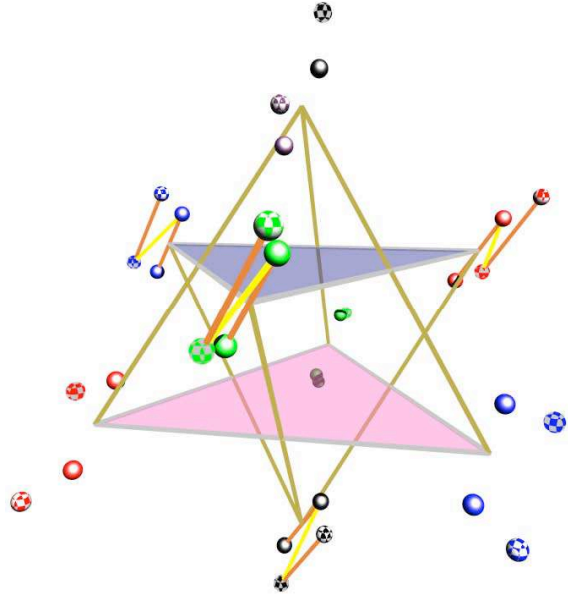}
\caption{Squarks and sleptons (checkered balls) added to the 
one-generation double simplex.\label{fig:susy}}
\end{center}
\end{figure}

Once the double simplex had taken shape, my colleagues and I amused 
ourselves by extending the basic object to display various sorts of 
physics beyond the standard model. I show a simple example in 
Figure~\ref{fig:susy}.
The superpartners of quarks and leptons are displayed as checkered
balls outboard of their standard-model partners.  Charged-current weak
interactions between squarks or slepons are mediated by the same
$W$-boson as those between quarks or leptons.  Depicting the
transitions by the same orange stalks in both cases is a metaphor for
the equal couplings of particles and their superpartners.  In addition,
supersymmetry implies new interactions that change quarks into squarks,
or leptons into sleptons.  These are mediated by winos, depicted by
yellow lines.  It is obvious how to represent strong interactions among
squarks, mediated by gluons, and quark--squark transitions, induced by
gluinos.

We have found it easy to construct visual metaphors for all manner of 
new physics, including technicolor, Ka\l uza--Klein excitations, and 
Majorana neutrinos. We have thus achieved the goal of devising an object 
that can be used not only to represent what we know, but also to 
illustrate new possibilities.

\section{Closing remarks \label{sec:close}}
The object of our double-simplex construction project has been to 
identify important topical questions for particle physics without 
plunging into formalism. As a theoretical physicist, I have deep 
respect for the power of mathematics to serve as a refiner's fire for 
our ideas. But I hope this exercise has helped you to see the power 
and scope of physical reasoning and the insights that can come from 
building and looking at a physical object with an inquiring 
spirit---even if the physical object inhabits an abstract space!

The mathematical underpinnings of the double simplex do bring discipline 
to the questions it elicits.
The structure of the double simplex is based on the
$\mathrm{{SU(4)}\otimes {SU(2)}\otimes {SU(2)}}$ decomposition of
$\mathrm{SO(10)}$.  A three-dimensional solid (tetrahedron) represents
the fundamental \textbf{4} representation of $\mathrm{SU(4)}$.  It is
decorated at the vertices with dumbbells representing the
{$\mathrm{SU(2)_{L}}$} and {$\mathrm{SU(2)_{R}}$} quantum numbers.  The
vertical coordinate of $\mathrm{SU(4)}$ can be read as $B-L$, the
difference of baryon number and lepton number.  The group
$\mathrm{SO(10)}$ is a useful classification symmetry, because its
16-dimensional fundamental representation contains an entire generation
of the known quarks and leptons.  Using $\mathrm{SO(10)}$ as a
coordinate system, if you like, carries no implication that it is the
symmetry of the world, or that it is the basis of a unified theory of
the strong, weak, and electromagnetic interactions.

We have succeeded in mapping current knowledge in a manner that evokes
important questions, but we shouldn't be lulled into thinking that we
have found all the important questions.  We don't know whether we
possess all the important clues.  When Mendele'ev created his periodic 
table of the elements, he knew nothing of the noble gases. His 
seven-column table (in modern notation) was enormously informative and 
stimulating, but an attempted to create a theory of Mendele'ev's chart 
and nothing else would presumably have failed, because an important 
puzzle piece was missing. Might we be missing a crucial piece of our 
puzzle?

Let us summarize some questions we have encountered in the course of
building up the double simplex.  $\Box$~Are quarks and leptons
elementary?  $\Box$~What is the relationship of quarks to leptons?
$\Box$~Are there right-handed weak interactions?  $\Box$~Are there new
quarks and leptons?  $\Box$~Are there new gauge interactions linking
quarks and leptons?  $\Box$~What is the relationship of left-handed \&
right-handed particles?  $\Box$~What is the nature of the right-handed
neutrino?  $\Box$~What is the nature of the mysterious new force that
hides electroweak symmetry?  $\Box$~Are there different kinds of
matter?  $\Box$~Are there new forces of a novel kind?  $\Box$~What do
generations mean?  How many?  $\Box$ Which quarks go with which leptons
$\Box$ Is there a family symmetry?  $\Box$~What makes a top quark a top
quark, and an electron an electron?  $\Box$~What is the (grand)
unifying symmetry?  What hides it?

What can we say about the ``homework problem'' I posed at the end of 
\S\ref{sec:gr}? First, it is clear that quarks and leptons would remain massless,
because mass terms are not permitted in our left-handed world if the
electroweak symmetry remains manifest.\footnote{I assume for this
discussion that all the trappings of the Higgs mechanism, including
Yukawa couplings for the fermions, are absent.} We have done nothing to
QCD, so that would still confine the (massless) color-triplet quarks
into color-singlet hadrons, with very little change in the masses of
those stable structures.  In particular, the nucleon mass would be
essentially unchanged, but the proton would outweigh the neutron
because the down quark now does not outweigh the up quark, and that
change will have its own consequences.

An interesting and slightly subtle point is that, even in the absence
of a Higgs mechanism, the electroweak symmetry is broken by QCD. As we
approach low energy in QCD, confinement occurs and the chiral symmetry
that treated the massless left-handed and right-handed quarks as
separate objects is broken.  The resulting communication between the
left-handed and right-handed worlds engenders a breaking of the
electroweak symmetry.  The trouble is that the scale of electroweak
symmetry breaking is measured by the pseudoscalar decay constant of the
pion, so the amount of mass acquired by the $W$ and $Z$ is set by
$f_{\pi}$, not by what we know to be the electroweak scale: it is off
by a factor of 2500.

But the fact is that the electroweak symmetry is broken, so the world
without a Higgs mechanism---but with strong-coupling QCD---is a world
in which the $\mathrm{SU}(2)_{\mathrm{L}}\otimes \mathrm{U}(1)_{Y}$
becomes $\mathrm{U}(1)_{\mathrm{em}}$.  Because the $W$ and $Z$ have
masses, the weak-isospin force, which we might have taken to be a
confining force in the absence of symmetry breaking, is not confining.
Beta decay is very rapid, because the gauge bosons are very light.  The
lightest nucleus is therefore one neutron; there is no hydrogen atom.
Analyses of what would happen to big-bang nucleosynthesis in this world
suggest that some light elements such as helium would be created.
Because the electron is massless, the Bohr radius of the atom is
infinite, so there is nothing we would recognize as an atom, there is
no chemistry as we know it, there are no stable composite structures
like the solids and liquids we know.

I invite you to explore this scenario in even greater detail.  To do
so is at least as challenging as trying to understand the world we do
live in!  The point is to see how very different the world would be,
if it were not for the mechanism of electroweak symmetry breaking whose
inner workings we intend to explore and understand in the next decade.  What
we are really trying to get at, when we look for the source of
electroweak symmetry breaking, is why we don't live in a world so
different, why we live in the world we do.  I think that's a glorious
question, one of the deepest questions that human beings have ever
tried to engage, and we will find the answer!

In closing, I wish Gustavo well in his new and venerable estate, and I
count on the Lisbon school to be in the thick of pursuing the questions
we have discussed today.  Fermi National Accelerator Laboratory is
operated by Universities Research Association Inc.\ under Contract No.\
DE-AC02-76CH03000 with the U.S.\ Department of Energy.  I am grateful
to Gui Rebelo and the organizing committee for the kind invitation to
participate in this celebration, and for warm hospitality.  I thank
many colleagues, notably Carl Albright, Uli Baur, Bogdan Dobrescu,
Chris Hill, Andreas Kronfeld, Joe Lykken, Jack Marburger, Uli Nierste,
Yasonuri Nomura, Dave Rainwater, and Maria Spiropulu, for their
valuable contributions to the development of double simplex.  My
primitive sketchbook, containing interactive graphics and photographs
of ball-and-stick models, with a minimal explanatory text, is available
for browsing at \url{http://lutece.fnal.gov/DoubleSimplex}.

\end{document}